\documentclass[12pt,epsfig]{article}
\usepackage{amssymb,amsmath, wasysym}
\usepackage{graphicx, float,psfrag}
\usepackage{bm}
\usepackage{color,colordvi}
\usepackage{feynarts}
\usepackage{epsfig}
\usepackage{axodraw}
\usepackage[numbers]{natbib}
\usepackage[]{graphicx,bm}
\usepackage{epsfig}

\newcommand{\lsim}{\lesssim}
\def\ga{\mathrel{\raise.3ex\hbox{$>$\kern-.75em\lower1ex\hbox{$\sim$}}}}
\def\la{\mathrel{\raise.3ex\hbox{$<$\kern-.75em\lower1ex\hbox{$\sim$}}}}

\begin{document}
\renewcommand{\theequation}{\arabic{section}.\arabic{equation}}

\title{\vspace{-50mm}
       \vspace{2.5cm}
       \vspace{8mm}
Higgs boson decay into 2 photons in the type~II Seesaw Model}
\vspace{2mm}
\author{A.~Arhrib${}^{1,2}$, R. Benbrik${}^{2,3, 4}$,  M.~Chabab${}^2$, 
        \\ G.~Moultaka${}^*$ ${}^{5,6}$ 
L.~Rahili${}^2$ \\[3mm]
{\normalsize \em ${}^1$  D\'epartement de Math\'ematiques, Facult\'e 
des Sciences et Techniques, Tanger, Morocco}\\
{\normalsize \em ${}^2$Laboratoire de Physique des Hautes Energies 
et Astrophysique } \\
{\normalsize \em  Universit\'e Cadi-Ayyad, FSSM, Marrakech, Morocco} \\
{\normalsize \em ${}^3$ Facult\'e Polydisciplinaire, Universit\'e 
Cadi Ayyad, Sidi Bouzid, Safi-Morocco}\\
{\normalsize \em ${}^4$ Instituto de Fisica de Cantabria (CSIC-UC), 
Santander, Spain} \\
{\normalsize \em ${}^5$ Universit\'e Montpellier 2, Laboratoire Charles 
Coulomb UMR 5221,}\\
{\normalsize \em F-34095 Montpellier, France}\\
{\normalsize \em ${}^6$ CNRS, Laboratoire Charles Coulomb UMR 5221,  F-34095
  Montpellier, France}
}
\setcounter{figure}{0}
\maketitle
\thispagestyle{empty}
\begin{abstract}

We study the two photon decay channel of the Standard Model-like
component of the CP-even Higgs bosons present in the type II Seesaw Model.
The corresponding cross-section is found to be significantly enhanced
in parts of the parameter space, due to the (doubly-)charged Higgs bosons'  $(H^{\pm \pm})H^\pm$
virtual contributions, while all the other Higgs decay channels remain Standard Model(SM)-like.
In other parts of the parameter space  $H^{\pm \pm}$ (and $H^{\pm}$) interfere destructively, 
reducing  the two photon branching ratio tremendously  below the SM prediction. 
Such properties allow to account for any excess such as the one reported by ATLAS/CMS
at $\approx 125$~GeV if confirmed by future data; if not, for the fact that a SM-like Higgs exclusion 
in the diphoton channel around $114$--$115$~GeV as reported by ATLAS, does not contradict a SM-like
Higgs at LEP(!), and at any rate, for the fact that ATLAS/CMS exclusion limits put
stringent lower bounds on the $H^{\pm \pm}$ mass, particularly in
the parameter space regions where the direct limits from 
same-sign leptonic decays of $H^{\pm \pm}$ do not apply.

\end{abstract}
{ ${}^*${\sl corresponding author}}
\newpage
\section{Introduction}
\label{sec:intro}
The LHC running at 7 TeV center of mass energy is accumulating more and more data. The ATLAS 
and CMS experiments have already probed
the Higgs boson in the mass range $110$-- $600$ GeV, and excluded a Standard Model (SM) Higgs  in 
the range $141$--$476$ GeV at the $95 \%$C.L. through a combined analysis of 
all decay channels and up to $\sim 2.3 {\rm fb}^{-1}$ integrated luminosity per experiment, 
\cite{ATLAS-CONF-2011-157-CMS-PAS-HIG-11-023}. 
Very recently, the analyses of $4.9 {\rm fb}^{-1}$ datasets 
for the combined channels made separately by ATLAS and by CMS,  have narrowed further down the mass window 
for a light SM Higgs, 
excluding respectively the mass ranges $131$--$453$ GeV (apart from the range $237$--$251$ GeV), 
\cite{ATLAS-CONF-2011-163}, and $127$--$600$ GeV  \cite{CMS-PAS-HIG-11-032} at the  $95 \%$C.L.  
More interestingly, both experiments exclude $1$ to $2$--$3$ times the SM diphoton cross-section 
at the  $95 \%$C.L. in
most of the mass range $110$--$130$~GeV, and report an excess of events
around $123$--$127$~GeV in the diphoton channel (as well as, but with lower statistical significance, 
in the $WW^*$ and $ZZ^*$ channels), corresponding to an exclusion of $3$ and $4$ times the SM cross-section respectively for CMS 
\cite{CMS-PAS-HIG-11-030} and ATLAS \cite{ATLAS-CONF-2011-161}. Furthermore, they exclude a SM Higgs in small, 
though different, portions of this mass range, $(112.7)114$--$115(.5)$~GeV for ATLAS and $127$--$131$ GeV for CMS, 
at the $95 \%$C.L. 

Notwithstanding the very exciting perspective of more data to come during the next LHC run, one remains  for the 
time being free to interpret the present results as either pointing towards a SM Higgs around $125$~GeV, or 
to a non-SM Higgs  around $125$~GeV in excess of  a few factors in the diphoton channel, or to behold
that these results are still compatible with statistical fluctuations.

The main purpose of the present paper is \underline{not} to show that the model we consider can account
for a Higgs with mass $\approx 125$GeV, although it can do so as will become apparent in the sequel.
Our aim will be rather to consider more globally how the recent experimental exclusion limits can constrain
 the peculiar features we will describe of the SM-like component of the model.

Although ATLAS/CMS exclusion limits assume SM-like branching ratios for all search channels,
they can also be used in case the branching ratio of {\sl only} the diphoton decay channel,
${\rm Br}(H \to \gamma \gamma)$, 
differs significantly from its SM value.
This is due to the tininess of this branching ratio ($\la 2\times 10^{-3}$), so that
if enhanced even  by more than an order of magnitude, due to the effects of some non-standard physics, 
all the other branching ratios would remain essentially unaffected. Thus, the present SM-like exclusion 
limits for the {\sl individual channels} could still be directly applied. Furthermore, if 
this non-standard physics keeps the tree-level Higgs couplings  to fermions and to W and Z gauge bosons very 
close to the SM ones, then obviously the corresponding channels will not lead to exclusions specific 
to this new physics.  
 The diphoton channel becomes then of particular interest in this case and can already constrain
parts of the parameter space of the new physics through the present exclusion limits
in the Higgs mass range $114$--$130$GeV.

A natural setting for such a scenario is the Higgs sector of the so-called Type II Seesaw Model for 
neutrino mass generation \cite{Konetschny:1977bn, Cheng:1980qt, Lazarides:1980nt, Schechter:1980gr, Mohapatra:1980yp}.
This sector, containing two CP-even, one CP-odd, one charged and one doubly-charged Higgs scalars,
 can be tested directly at the LHC, provided that
the Higgs triplet mass scale $M_{\Delta}$ and the soft lepton-number violating mass parameter 
$\mu$ are of order or below the weak-scale  \cite{Akeroyd:2007zv, Perez:2008ha, delAguila:2008cj,  
Akeroyd:2009nu, Fukuyama:2009xk,Akeroyd:2009hb,Petcov:2009zr, Fukuyama:2010mz,
Akeroyd:2010ip, Arhrib:2011uy}.  Moreover, in most of the parameter space  [and apart from an extremely narrow 
region of $\mu$], one of the two  CP-even Higgs scalars is generically essentially SM-like and the other an almost 
decoupled triplet, irrespective of their relative masses, \cite{Arhrib:2011uy}. It follows that 
if all the Higgs sector of the model is accessible to the LHC, one expects a neutral Higgs state with
cross-sections very close to the SM in all Higgs production and decay channels to leading electroweak order,
except for the diphoton (and also $\gamma Z$) channel. Indeed, in the latter channel, loop effects of the other 
Higgs states can lead to substantial enhancements which can then be readily analyzed in the light of the experimental
exclusion limits as argued above. 

In this paper we will study quantitatively this issue. The main result is that
the loop effects of the charged and in particular the  doubly-charged Higgs states can either enhance
the diphoton cross-section by several factors, or reduce it in some cases by several orders
of magnitude essentially without affecting the other SM-like decay channels.  This is consistent with the present 
experimental limits on these (doubly-)charged Higgs states masses and can be interpreted in several ways. 
It can account for an excess in the diphoton cross-section like the one observed by ATLAS/CMS.
But it can also account for a deficit in the diphoton cross-section without affecting the other
channels. The latter case could be particularly interesting for the $114$--$115$~GeV SM Higgs mass range excluded 
by ATLAS, [provided one is willing to interpret the excess at $\approx 125$GeV as statistical fluctuation]. Indeed,
since the coupling of the Higgs to the $Z$ boson remains standard in our model, a possible LEP signal
at  $114$--$115$~GeV would remain perfectly compatible with the ATLAS exclusion!

The rest of the paper is organized as follows: in section 2 we briefly review some
ingredients of the Higgs sector of the type II seesaw model, hereafter dubbed DTHM.
In section 3 we calculate the branching ratio of
$H\to \gamma\gamma$ in the context of DTHM and discuss its sensitivity
to the parameters of the model.[The $\gamma Z$ channel can be treated along similar
lines but will not be discussed in the present paper.]
Section 4 is devoted to the theoretical and experimental constraints as well as
to the numerical analysis for the physical observables.
We conclude in section 5.  
 
\section{The DTHM Model}
\label{sec:The DTHM model}

In \cite{Arhrib:2011uy} we have performed a detailed study of DTHM potential,
derived the most general set of dynamical constraints on the parameters of the
model at leading order and outlined the salient features of Higgs boson phenomenology at the colliders. 
These constraints delineate precisely the theoretically allowed parameter 
space domain that one should take into account in Higgs phenomenological 
analyses. We have also shown that in most of the parameter space the DTHM
 is similar to the SM except in the small $\mu$ regime where the doublet and
 triplet component of the Higgs could have a maximal mixing. \\

The scalar sector of the DTHM model consists of the 
standard Higgs doublet $H$ and a colorless Higgs triplet $\Delta$ with
hypercharge $Y_H=1$ and $Y_\Delta=2$ respectively. 
Their matrix representation are given by:
\begin{eqnarray}
\Delta &=\left(
\begin{array}{cc}
\delta^+/\sqrt{2} & \delta^{++} \\
\delta^0 & -\delta^+/\sqrt{2}\\
\end{array}
\right) \qquad {\rm and} \qquad H=\left(
                    \begin{array}{c}
                      \phi^+ \\
                      \phi^0 \\
                    \end{array}
                  \right)
\end{eqnarray}
The most general $SU(2)_{L}\times U(1)_{Y}$ gauge invariant 
renormalizable Lagrangian in the scalar sector is 
\cite{Perez:2008ha,Arhrib:2011uy}:

\begin{eqnarray}
\mathcal{L} &=&
(D_\mu{H})^\dagger(D^\mu{H})+Tr(D_\mu{\Delta})^\dagger(D^\mu{\Delta}) 
-V(H, \Delta) + \mathcal{L}_{\rm Yukawa}
\label{eq:DTHM}
\end{eqnarray}

where the potential $V(H, \Delta)$ is given by,
\begin{eqnarray}
V&=&-m_H^2{H^\dagger{H}}+\frac{\lambda}{4}(H^\dagger{H})^2+
M_\Delta^2Tr(\Delta^{\dagger}{\Delta}) + 
\lambda_1(H^\dagger{H})Tr(\Delta^{\dagger}{\Delta})\label{eq:VDTHM} \\ & + & 
\lambda_2(Tr\Delta^{\dagger}{\Delta})^2
+\lambda_3Tr(\Delta^{\dagger}{\Delta})^2 
+ \lambda_4{H^\dagger\Delta\Delta^{\dagger}H}+
[\mu(H^T{i}\tau_2\Delta^{\dagger}H)+hc]\nonumber
\end{eqnarray}
${\mathcal{L}}_{Yukawa}$ contains all the SM Yukawa sector plus 
one extra term that provides, after spontaneous electroweak symmetry 
breaking (EWSB), a Majorana mass to neutrinos.

Once EWSB takes place, the Higgs doublet and triplet acquire vacuum 
expectation values
\begin{equation}
\langle H \rangle = \frac{1}{\sqrt{2}} \left(
                    \begin{array}{c}
                      0 \\
                      v_d \\
                    \end{array}
                  \right), \qquad \langle \Delta \rangle = \frac{1}{\sqrt{2}}
\left(
\begin{array}{cc}
0 & 0 \\
v_t & 0\\
\end{array}
\right)
\label{vacuum}
\end{equation}
inducing the Z and W masses
\begin{equation}
M_Z^2=\frac{(g^2+{g'}^2)(v_d^2+4v_t^2)}{4} \quad , \quad\\
M_W^2=\frac{g^2(v_d^2+2v_t^2)}{4}  \label{eq:mW}
\end{equation}
with $v^2=(v_d^2+4v_t^2)\approx(246~ \mathrm{GeV})^2$.

The DTHM is fully specified by seven independent parameters which we will
take: $\lambda$, $\lambda_{i=1...4}$, $\mu$  and $v_t$. 
These parameters respect a set of dynamical 
constraints originating from the potential , particularly 
perturbative unitarity and boundedness from below constraints . \\ 
The model spectrum contains seven physical Higgs states: 
a pair of CP even states $(h^0, H^0)$, one CP odd Higgs boson $A$, 
one simply charged Higgs $H^\pm$ and one doubly charged state $H^{\pm\pm}$.
The squared masses of the neutral CP-even states and of the charged and doubly charged states
are given in terms of the VEV's and the parameters of the potential as follows,

\begin{eqnarray}
m_{h^0}^2&=&\frac{1}{2}[A+C-\sqrt{(A-C)^2+4B^2}] \label{eq:mh0}\\
m_{H^0}^2&=&\frac{1}{2}[A+C+\sqrt{(A-C)^2+4B^2}] \label{eq:mH0}
\end{eqnarray}
with
\begin{eqnarray}
  A &=& \frac{\lambda}{2}{v_d^2}, \;
  B =v_d ( -\sqrt{2}\mu+(\lambda_1+\lambda_4)v_t) , \; 
  C = \frac{\sqrt{2}\mu{v_d^2}+4(\lambda_2+\lambda_3)v_t^3}{2v_t} \label{eq:ABC} \nonumber
\end{eqnarray}
and 
\begin{eqnarray}
m_{H^\pm}^2&=&\frac{(v_d^2+2 v_t^2)[2\sqrt{2}\mu- \lambda_4 v_t]}{4v_t}
\label{eq:mHpm} \\
m_{H^{\pm\pm}}^2&=&\frac{\sqrt{2}\mu{v_d^2}- \lambda_4v_d^2v_t-2\lambda_3v_t^3}{2v_t}  \label{eq:mHpmpm}
\end{eqnarray} 
For a recent and comprehensive study of the DTHM, in particular concerning the distinctive properties of the mixing 
angle between the neutral components of the doublet and triplet  Higgs fields, we refer the reader to \cite
{Arhrib:2011uy}.

We close this section by stressing an important point which is seldom clearly stated in
the literature. Recall first that the general rational justifying the name 'type II seesaw' assumes 
$\mu \sim M_\Delta \sim M_{\rm GUT}$ (or any other scale much larger than the electroweak scale). 
One then obtains naturally $v_t \ll v_d$, as a consequence of the electroweak symmetry breaking conditions,
and thus naturally very small neutrino masses for Yukawa couplings of order 1. 
But then one has also $\mu \gg v_t$ and consequently a very heavy Higgs sector, largely out of 
the reach of the LHC, apart from the lightest state $h^0$, as can be seen from the above mass expressions; 
this leaves us at the electroweak scale with simply a SM Higgs sector. 
Put differently, a search for the DTHM Higgs states at the LHC entails small $\mu (\sim {\cal O}(v_t))$
and thus implicitly questions the validity of the seesaw mechanism. 
Since we are interested in new physics visible at the LHC, we will take up the latter assumption of 
small $\mu$ in our phenomenological study, which can also have some theoretical justification related 
to spontaneous soft lepton-number violation.

\section{$\mathcal{H}\to \gamma\gamma$}
\label{sec:Higgs decay}
The low SM Higgs mass region, $[110,140]$ GeV, is the most challenging for LHC searches.
In this mass regime, the main search channel through the rare decay into a pair of photons
can be complemented by the decay into $\tau^+ \tau^-$
and potentially the $b \bar{b}$ channel (particularly for the lower edge of the mass range
and/or for supersymmetric Higgs searches), while the $WW^*, ZZ^*$ channels 
are already competitive in the upper edge ($130$--$140$ GeV) of this mass range 
\cite{ATLAS-CONF-2011-157-CMS-PAS-HIG-11-023} 
the Higgs being produced mainly via gluon fusion \cite{Aad:2011qi}, \cite{Ball:2007zza}.
  
The theoretical predictions for the loop induced decays $H\to \gamma\gamma$ 
(and $H \to \gamma Z$) have been initiated since many 
years \cite{Ellis:1975ap, Ioffe:1976sd, Shifman:1979eb}.
Several more recent studies have been carried out looking for 
large loop effects. 
Such large effects can be found in various extensions of the SM, such as the Minimal
Supersymmetric Standard Model (MSSM) 
\cite{Djouadi:1996pb,Djouadi:1996yq,Huang:2001tj,Carena:2011mw}, the two Higgs Doublet
Model \cite{Ginzburg:2001wj,Arhrib:2004ak,Posch:2010hx,LopezVal:2011ak,Burdman:2011ki},
the Next-to-MSSM \cite{Ellwanger:2010nf,Moretti:2006sv, Ellwanger:2011aa}, the little Higgs models 
\cite{Han:2003gf,Wang:2011rv} and
 in models with a real triplet \cite{FileviezPerez:2008bj}. 
To the best of our knowledge there is no $H \to \gamma\gamma$ study in the context
of a triplet field with hypercharge $Y=2$, that is comprising charged and doubly-charged
Higgs states.

We turn here to the study of the latter case explaining how these charged and doubly charged
Higgs states of the DTHM 
could enhance or suppress the 2 photons decay rate.
Furthermore, since one or the other of the two CP-even neutral Higgs bosons $h^0, H^0$ present in the DTHM
can behave as a purely SM-like Higgs depending on the regime under consideration (see
\cite{Arhrib:2011uy}), we will refer to the SM-like state generically as $\mathcal{H}$ in the
following. It should be kept in mind, however, that when  $\mathcal{H}= h^0$ all the other
DTHM Higgs states are heavier than $\mathcal{H}$ while when $\mathcal{H}=H^0$ they are all
generically 
lighter than $\mathcal{H}$, thereby leading possibly to a different phenomenological interpretation
of the present experimental exclusion limits for $\mathcal{H}\to \gamma\gamma$ channel.

\begin{figure}[!ht]
\label{fig0}
\centering
\includegraphics[width=5in]{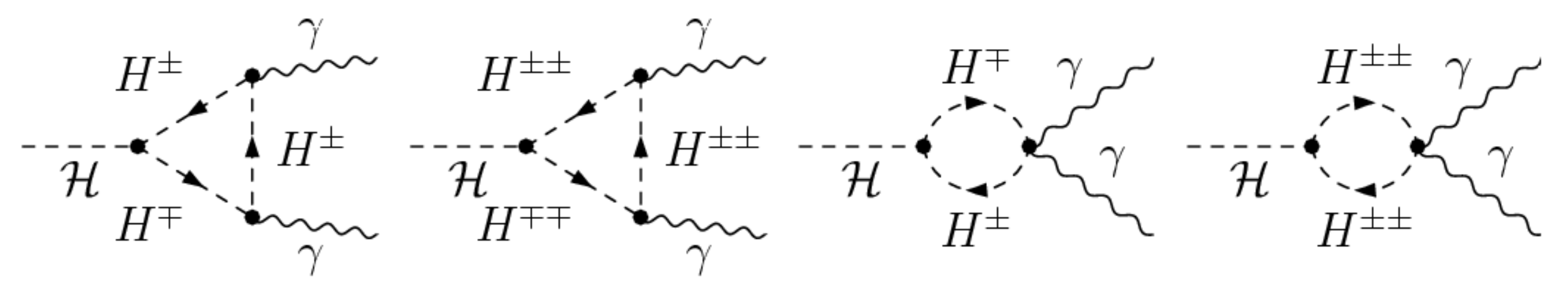}
\caption{Singly and doubly charged Higgs bosons contributions to
$\mathcal{H}\,(h^0,H^0)\to \gamma\gamma$ in the DTHM.}
\end{figure}

The decay $\mathcal{H}\to \gamma\gamma$ is mediated at 1-loop level by the virtual exchange of
the SM fermions, the SM gauge bosons and the new charged Higgs states. 
Using the general results for spin-$1/2$, spin-$1$ and spin-$0$ contributions, \cite{Shifman:1979eb}
(see also \cite{Gunion:1989we}, \cite{Spira:1997dg}, \cite{Djouadi:2005gj}),
one includes readily the extra  contributions to the partial width which takes the following form, 
\begin{eqnarray}
\label{eq:DTHM-h2gaga}
\Gamma({\mathcal{H}} \rightarrow\gamma\gamma)
& = & \frac{G_F\alpha^2 M_{{\mathcal{H}}}^3}
{128\sqrt{2}\pi^3} \bigg| \sum_f N_c Q_f^2 \tilde{g}_{{\mathcal{H}} ff} 
A_{1/2}^{{\mathcal{H}}}
(\tau_f) + \tilde{g}_{{\mathcal{H}} WW} A_1^{{\mathcal{H}}} (\tau_W) \nonumber \\
&& + \tilde{g}_{\mathcal{H} H^\pm\,H^\mp}
A_0^{{\mathcal{H}}}(\tau_{H^{\pm}})+
 4 \tilde{g}_{\mathcal{H} H^{\pm\pm}H^{\mp\mp}}
A_0^{{\mathcal{H}}}(\tau_{H^{\pm\pm}}) \bigg|^2
\label{partial_width_htm}
\end{eqnarray}
where the first two terms in the squared amplitude are the known SM
contributions up to the difference in the couplings of $\mathcal{H}$ to up and down quarks
and $W^\pm$ in the DTHM, when $\mathcal{H}$ is not purely SM-like.  
The relevant reduced couplings (relative to the SM ones) are summarized in Table.~\ref{table_couplings}.
In Eq.~(\ref{partial_width_htm}) $N_c=3 (1)$ for quarks (leptons), $Q_f$ is the 
electric charge of the SM fermion $f$. 
The scalar functions $A_{1/2}^{{\mathcal{H}}}$ for fermions and 
$A_{1}^{{\mathcal{H}}}$ for gauge bosons are known in the literature
and will not be repeated here (for a review see for instance \cite{Djouadi:2005gj}).
The last two terms correspond to the $H^\pm$ and $H^{\pm \pm}$ contributions
whose Feynman diagrams are depicted in Fig~\ref{fig0}. 
The structure of the $H^\pm$ and $H^{\pm \pm}$
contributions is the same except for the fact that the  $H^{\pm \pm}$
contribution is enhanced by a relative factor four in the amplitude
since $H^{\pm  \pm}$ has an electric charge of $\pm 2$ units. 
The scalar function for spin-$0$ 
 $A_0^{{\mathcal{H}}}$ is defined as 

\begin{eqnarray}
A_{0}^{{\mathcal{H}}}(\tau) &=& -[\tau -f(\tau)]\, \tau^{-2}
\label{eq:Ascalar}
\end{eqnarray}
with $\tau_{i}=m^2_{\mathcal{H}}/4m^2_{i}$ $(i=f,W,H^{\pm},H^{\pm\pm})$ 
and the function $f(\tau)$ is given by
\begin{eqnarray}
f(\tau)=\left\{
\begin{array}{ll}  \displaystyle
\arcsin^2\sqrt{\tau} & \tau\leq 1 \\
\displaystyle -\frac{1}{4}\left[ \log\frac{1+\sqrt{1-\tau^{-1}}}
{1-\sqrt{1-\tau^{-1}}}-i\pi \right]^2 \hspace{0.5cm} & \tau>1
\end{array} \right. 
\label{eq:ftau} 
\end{eqnarray}
while the {\sl reduced} DTHM trilinear couplings of $\mathcal{H}$ to $H^\pm$ and $H^{\pm \pm}$ 
are given by

\begin{eqnarray}
\tilde{g}_{\mathcal{H} H^{++}H^{--}}  & = & - \frac{s_W}{e} \frac{m_W}{m_{H^{\pm \pm}}^2} g_{\mathcal{H} H^{++}H^{--}}
 \label{eq:redgcalHHpp}\\
\tilde{g}_{\mathcal{H} H^+H^-} & = & - \frac{s_W}{e} \frac{m_W}{m_{H^{\pm}}^2} g_{\mathcal{H} H^+H^-} \label{eq:redgcalHHp}
\end{eqnarray}
with
\begin{eqnarray}
g_{\mathcal{H} H^{++}H^{--}}  &\approx &  - \bar{\epsilon} \lambda_1v_d \label{eq:gcalHHpp}\\
g_{\mathcal{H} H^+H^-} & \approx & - \bar{\epsilon} (\lambda_{1} + \frac{\lambda_{4}}{2}) v_d \label{eq:gcalHHp}
\end{eqnarray}
The latter can be read off from the couplings of $h^0$, 

\begin{eqnarray}
g_{h^0H^{++}H^{--}} &=&-\{2\lambda_2v_ts_\alpha+\lambda_1v_dc_\alpha\} 
 \label{eq:ghHpp}\\
g_{h^0H^+H^-}&=&-\frac{1}{2}
\bigg\{\{4v_t(\lambda_2 + \lambda_3) c_{\beta'}^2+2v_t\lambda_1s_{\beta'}^2-
\sqrt{2}\lambda_4v_dc_{\beta'}s_{\beta'}\}s_\alpha  \\
&&+\{\lambda\,v_ds_{\beta'}^2+{(2\lambda_{1}+\lambda_{4}) }v_dc_{\beta'}^2+
(4\mu-\sqrt{2}\lambda_4v_t)c_{\beta'}s_{\beta'}\}c_\alpha\bigg\}
\nonumber  
\label{eq:ghHp}
\end{eqnarray}
in the limit where $h^0$ becomes a pure SM Higgs, 
i.e. when $s_\alpha \to 0$, or from the couplings of $H^0$, 
obtained simply from the above couplings by the substitutions 
\begin{eqnarray}
g_{H^0H^{++}H^{--}}&=&  g_{h^0H^{++}H^{--}}  [c_\alpha \rightarrow -s_\alpha, s_\alpha \rightarrow c_\alpha]
\label{eq:gHHpp}\\
g_{H^0H^+H^-}&=& g_{h^0H^+H^-}  [c_\alpha \rightarrow -s_\alpha, s_\alpha \rightarrow c_\alpha]
\label{eq:gHHp} 
\end{eqnarray}
in the limit where $H^0$ becomes a pure SM Higgs, 
i.e. when $c_\alpha \to 0$,
taking also into account that $ s_{\beta'}\approx  \sqrt{2} v_t/v_d$ with 
 $v_t/v_d \ll 1$. 
[In the above equations $\alpha$ and  $\beta'$ stand for the mixing angles in
 the  CP-even and charged Higgs sectors with the shorthand notations $s_x, c_x$ for $\cos x, \sin x$;
In Eqs.~(\ref{eq:gcalHHpp}, \ref{eq:gcalHHp}) we have denoted by $\bar{\epsilon}$ the sign of 
$s_\alpha$ in the convention where $c_\alpha$ is always positive, which is defined as
$\bar{\epsilon} = 1$ for $\mathcal{H} \equiv h^0$ and 
$\bar{\epsilon} = {\rm sign}[ \sqrt{2}\mu - (\lambda_1+\lambda_4) v_t]$ for $\mathcal{H} \equiv H^0$;
see \cite{Arhrib:2011uy}.] 
Obviously, in the limit where one of the two CP-even Higgs states is SM-like,
the other state behaves as a pure triplet $\Delta^0$ with suppressed couplings to $H^\pm$ and $H^{\pm\pm}$
given by  $g_{\Delta^0H^+H^-} \approx (\lambda_4/\sqrt{2}-2 (\lambda_2 + \lambda_3)) v_t$
and $g_{\Delta^0H^{++}H^{--}} \approx -2\lambda_2 v_t$. Due to the
smallness of $v_t/v_d$ the states $h^0, H^0$ are mutually essentially SM-like or essentially triplet,
apart from a very tiny and fine-tuned region where they carry significant components of both.
(see  \cite{Arhrib:2011uy} for more details).
We can thus safely consider that any experimental limit on the SM Higgs decay in
two photons can be applied exclusively either to $h^0$ or to $H^0$, depending on whether 
we assume  $\mathcal{H}$ to be the lightest or the heaviest among all the neutral and charged Higgs states
of the DTHM.\footnote{The above mentioned tiny region with mixed states can also be treated, provided one includes 
properly the contribution of both $h^0$ and $H^0$, which are in this case almost degenerate in mass as well
as with all the other Higgs masses of the DTHM.}

As a cross-check on our tools, an independent calculation using the FeynArts and
FormCalc \cite{Hahn:2000kx,Hahn:1998yk} packages for which we provided a  DTHM model file 
was also carried out and we found perfect agreement with Eq.~(\ref{partial_width_htm}).
\begin{table}[!h]
\begin{center}
\renewcommand{\arraystretch}{1.5}
\begin{tabular}{|c|c|c|c|c|} \hline\hline
\ \ $\mathcal{H}$ \ \  &$\tilde{g}_{\mathcal{H} \bar{u}u}$&
                  $\tilde{g}_{\mathcal{H} \bar{d}d}$&
$\tilde{g}_{\mathcal{H} W^+W^-} $ \\ \hline\hline
$h^0$  & \ $\; c_\alpha/c_{\beta'} \; $ \
     & \ $ \; c_\alpha/c_{\beta'} \; $ \
     & \ $ \; +e(c_\alpha\,v_d+2s_\alpha\,v_t)/(2s_W\,m_W) \; $ \ \\
$H^0$  & \ $\; - s_\alpha/c_{\beta'} \; $ \
     & \ $ \; - s_\alpha/ c_{\beta'} \; $ \
     & \ $ \; -e(s_\alpha\,v_d-2c_\alpha\,v_t)/(2s_W\,m_W) \; $ \ \\ \hline\hline
\end{tabular}
\end{center}
\caption{ The CP-even neutral Higgs couplings to 
fermions and gauge bosons in the DTHM {\sl relative} to the SM Higgs couplings,
$\alpha$ and $\beta'$ denote the mixing angles respectively in the CP-even and
charged Higgs sectors, $e$ is the electron charge, $m_W$ the $W$ gauge boson
mass and $s_W$ the weak mixing angle.}
\label{table_couplings}
\end{table}
Clearly the contribution of the $H^{\pm\pm}$ and $H^{\pm}$ loops depends
on the details of the scalar potential.
The phase space function $A_0$ involves the scalar masses 
$m_{H^\pm}$ and $m_{H^{\pm\pm}}$, while $g_{\mathcal{H}
  H^+H^-}$ and $g_{\mathcal{H} H^{++}H^{--}}$ are functions of 
several Higgs potential parameters. It is clear from  
Eqs.~(\ref{eq:gcalHHpp}, \ref{eq:gcalHHp}) that those couplings are not suppressed
in the small $v_t$ and/or $\sin\alpha$ limit but have a contribution which is
proportional to the vacuum expectation value of the doublet field
and hence can be a source  of large enhancement of $\mathcal{H}\to \gamma \gamma$ (and
$\mathcal{H}\to \gamma Z$).

As well known, the decay width of 
$\mathcal{H} \to \gamma\gamma$ in the SM is dominated by the W loops which can
also interfere destructively with the subdominant top contribution.
In the DTHM, the signs of the couplings $g_{\mathcal{H}H^+H^-}$ and $g_{\mathcal{H} H^{++}H^{--}}$,
and thus those of the  $H^\pm$ and $H^{\pm \pm}$ contributions to $\Gamma({\mathcal{H}} \rightarrow\gamma\gamma)$,
are fixed respectively by the signs of $2\lambda_1+\lambda_4$ and $\lambda_1$,
Eqs.(\ref{eq:redgcalHHpp}, \ref{eq:redgcalHHp}, \ref{eq:gcalHHpp}, \ref{eq:gcalHHp}). 
However, the combined perturbative unitarity and 
potential boundedness from below (BFB) constraints derived in \cite{Arhrib:2011uy}
confine $\lambda_1, \lambda_4$ to small regions. For instance, in the case of vanishing
$\lambda_{2,3}$, $\lambda_1$ is forced to be positive while $\lambda_4$ can have either signs
but still with bounded values of $|\lambda_4|$ and $|2\lambda_1+\lambda_4|$. Moreover, since
we are considering scenarios where $\mu \sim {\cal O}(v_t)$, negative values of $\lambda_4$ can be
favored by the experimental bounds on the (doubly)charged Higgs masses, Eqs.~(\ref{eq:mHpm},
\ref{eq:mHpmpm}). 
For definiteness we stick in the following to $\lambda_1 >0$,  
although the sign of $\lambda_1$ can be relaxed if $\lambda_{2,3}$ are non-vanishing. 
Also in the considered mass range 
for $\mathcal{H}, H^\pm$ and $H^{\pm \pm}$ the function $A_{0}^{{\mathcal{H}}}(\tau)$
is real-valued and 
takes positive values in the range $ 0.3 - 1$. An increasing value of $\lambda_1$
will thus lead to contributions of $H^\pm$ and $H^{\pm \pm}$ that are constructive
among each other but destructive with respect to the sum of $W$ boson and top quark contributions. 
[Recall that ${\cal R}e A_{1}^{{\mathcal{H}}}(\tau)$ takes negative values in the range
$-12$ to $-7$.]
As we will see in the next section, this can either reduce tremendously the
branching ratio into diphotons, or increase it by an amount that can be already constrained
by the present ATLAS/CMS results.

\section{Theoretical and experimental constraints, Numerics and Discussions}
\label{sec:numerics}
In this section we present the theoretical and experimental 
constraints we will take into account, and illustrate our numerical results.

Besides the branching ratio of $\mathcal{H}\to \gamma\gamma$, we will consider 
the following observable:
\begin{equation}
R_{\gamma\gamma}(\mathcal{H})=\frac{(\Gamma(\mathcal{H}\rightarrow gg)
\times {\rm Br}(\mathcal{H}\rightarrow \gamma\gamma))^{DTHM}}
{(\Gamma(\mathcal{H}\rightarrow gg)\times 
{\rm Br}(\mathcal{H} \rightarrow \gamma\gamma))^{SM}}
\label{eq:Rgg}
\end{equation}
which can be viewed as an estimate of the ratio of DTHM to SM of the  gluon fusion 
Higgs production cross section  with a Higgs decaying into a photon pair.
One should, however, keep in mind the involved approximations: assuming only one intermediate
(Higgs) state, one should take the ratio of the parton-level cross-sections 
$\sigma( g g \rightarrow \gamma \gamma)$ in both models, which are given
by $ {\rm Br}(\mathcal{H} \rightarrow g g) \times {\rm Br}(\mathcal{H} \rightarrow \gamma\gamma)$. Using instead the
ratio $R_{\gamma\gamma}$ as defined in Eq.~(\ref{eq:Rgg}) 
relies on the fact that in the SM-like Higgs regime of DTHM, the branching ratios of all Higgs decay channels 
are the {\sl same}
as in the SM, except for  $\mathcal{H}\rightarrow \gamma\gamma$ (and $\mathcal{H}\rightarrow \gamma Z$)
where they can significantly differ but remain very small compared to the other 
decay channels, so that 
$\Gamma(\mathcal{H} \rightarrow {\rm all})^{DTHM}/\Gamma(\mathcal{H} \rightarrow {\rm all})^{SM} \approx 1$.  
A ratio such as $R_{\gamma\gamma}$ has the advantage  that all the leading QCD corrections
as well as PDF uncertainties drop out. There will be, however, other approximations involved
when identifying $R_{\gamma\gamma}$ with the ratio $\sigma^{\gamma \gamma}/\sigma_{SM}^{\gamma \gamma}$
that is constrained by the recent ATLAS and CMS limits, where  $\sigma^{\gamma \gamma} \equiv
\sigma^{\mathcal{H}} \times {\rm Br}(\mathcal{H} \to \gamma \gamma)$ and $\sigma^{\mathcal{H}}$
denotes the Higgs production cross-section. For instance we do not include all known QCD corrections
(see however section 4.3) and neglect the vector boson fusion Higgs production contribution
in our analysis.
 

\subsection{DTHM parameter scans and theoretical constraints}
All the Higgs mass spectrum of the model is fixed in terms
of $\lambda$, $\lambda_{1,2,3,4}$,
$v_t$ and $\mu$ which we will take as input parameters, \cite{Arhrib:2011uy}.
As one can see from Eq.~(\ref{eq:VDTHM}) $\lambda_{2}$ and $\lambda_3$ 
enter only the purely triplet sector. Since we focus here on the SM-like (doublet) component, 
their contributions will  always be suppressed by the triplet VEV value and can be safely neglected
as compared to the contributions of  $\lambda_{1}$ and $\lambda_4$ which enter the game 
associated with the doublet VEV,  Eqs.~(\ref{eq:ghHpp} - \ref{eq:gHHp}).  
Taking into account the previous comments, $\lambda_{2,3}$ will be fixed and we perform a 
scan over the other parameter as follows: 
\begin{center}
${\it v}_t=1\,{\rm GeV}$\,\qquad $\lambda = 0.45 \sim 1$\,
\qquad $0 < \lambda_1 < 10$\\
$\lambda_3=2\lambda_2 = 0.2$\,\qquad  $0.2 < \mu < 20$\,
\qquad $-5 < \lambda_4 < 3$
\end{center}
The chosen range for $\lambda$ values ensures a light SM-like Higgs state
and the scanned domain of the $\lambda_i$'s is consistent with
the perturbative unitarity and BFB bounds mentioned earlier.

\subsection{Experimental constraints}
Here we will discuss the experimental constraints on 
the triplet vev as well as on the scalar particles of the DTHM.
In the above scan, the triplet vev has been taken equal to 1 GeV in order to satisfy 
the constraint from the $\rho$ parameter \cite{Amsler:2008zzb} for which the 
tree-level extra contribution $\delta\rho$ should not exceed the current limits
from precision measurements: $|\delta\rho| \lesssim 0.001$.\\
Nowadays, the doubly charged Higgs boson is subject to many experimental
searches. Due to its  spectacular signature 
from $H^{\pm \pm}\to l^\pm l^\pm$, the doubly charged Higgs has been
searched by many experiments such as LEP, Tevatron and LHC.
At the Tevatron, D$\emptyset$\cite{Abazov:2004au},\cite{:2008iy} 
and CDF \cite{Acosta:2004uj}, \cite{Aaltonen:2008ip}
 excluded a doubly charged Higgs with a mass
in the range $100\to 150$ GeV. Recently, CMS also performed with 1 fb$^{-1}$ luminosity 
a search for doubly charged Higgs decaying to a pair of leptons,
setting  a lower mass limit of 313 GeV from 
$H^{\pm\pm}\to \mu^\pm\mu^\pm, e^\pm e^\pm,
\mu^\pm e^\pm $ \cite{CMS-PAS-HIG-11-010}.
The limit is lower if we consider the other decay channel with 
one electron or more \cite{CMS-PAS-HIG-11-001}, \cite{CMS-PAS-HIG-11-010}.

We stress that all those bounds assume
 a 100\% branching ratio for $H^{\pm\pm}\to l^\pm l^\pm$ decay, while in
 realistic cases one can easily find scenarios where this decay
channel is suppressed whith respect to 
$H^{\pm\pm}\to W^{\pm} W^{\pm (*)}$  
\cite{Perez:2008ha,Akeroyd:2007zv,Garayoa:2007fw, Kadastik:2007yd}
which could invalidate partially the CDF, D$\emptyset$, CMS and
ATLAS limits. In our scenario with $v_t \lsim 1$~GeV we estimated
that the decay channel $H^{\pm\pm}\to W^{\pm} W^{\pm *}$ can still overwhelm
the two-lepton channel for $m_{H^{\pm\pm}}$ down to $\approx 110 GeV$.
We will thus take this value as a nominal lower bound in our numerical analysis.

In the case of charged Higgs boson, 
if it decays dominantly to leptons or to light quarks $cs$ (for small $v_t$) 
we can apply the LEP mass lower bounds that are of the order of 80 GeV 
\cite{Achard:2003gt}, \cite{Abdallah:2003wd}.
For large $v_t$, i.e. much larger than the neutrino masses but 
still well below the electroweak scale,  the dominant decay is 
either $H^+\to t\bar{b}$ or one of the
bosonic decays $H^+\to W^+Z$, $H^+\to W^+h^0/W^+A^0$ . For the first two decay modes there has been no
explicit search neither at LEP nor at the Tevatron, while for the $H^+\to W^+A^0$ decay (and possibly 
for $H^+\to W^+ h^0/$ if $h^0$ decays similarly to $A^0$), one can use the LEPII search performed
 in the framework of two Higgs doublet models. In this case the charged Higgs mass limit is again of the order 
of 80 GeV \cite{Abdallah:2003wd}.


\subsection{Numerical results}

In the subsequent numerical discussion we use the following input parameters: 
$G_F= 1.166 \times 10^{-5}$ GeV$^{-2}$, $\alpha^{-1} \approx 128$, $m_Z=91.1875$G~eV, 
$m_W=80.45$~GeV and $m_t=173$~GeV. 
We also compute the total width of the Higgs boson  taking into account leading order  QCD 
corrections as given in \cite{Djouadi:1995gv} as well as the 
off-shell decays $\mathcal{H}\to WW^*$ and $\mathcal{H}\to ZZ^*$
\cite{Rizzo:1980gz, Keung:1984hn}.

\begin{figure}[!h]
\hspace{-1.1cm}
  \begin{tabular}{cc}
    \resizebox{72mm}{!}{\includegraphics[scale=0.9]{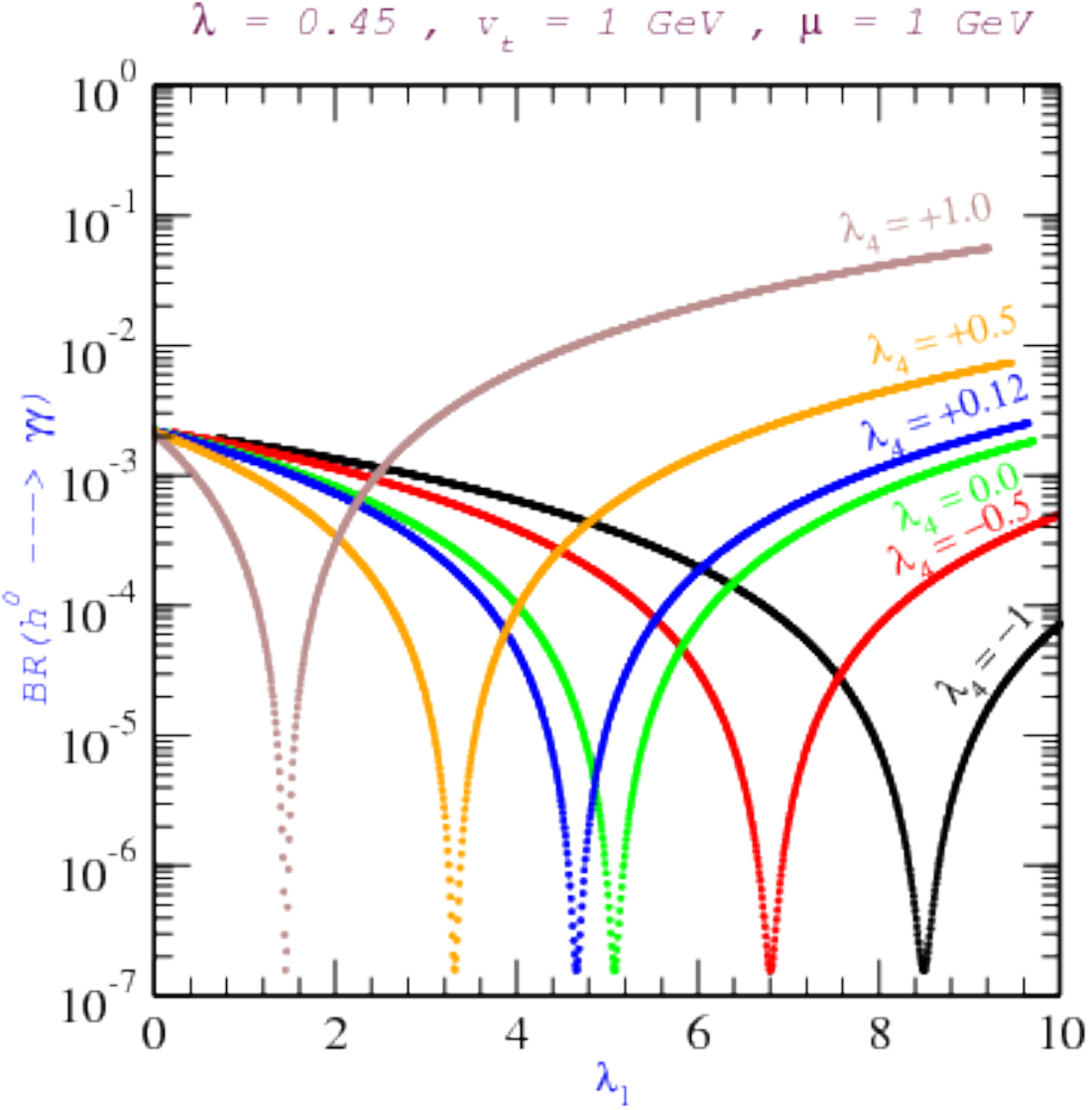}} &
    \resizebox{72mm}{!}{\includegraphics[scale=0.9]{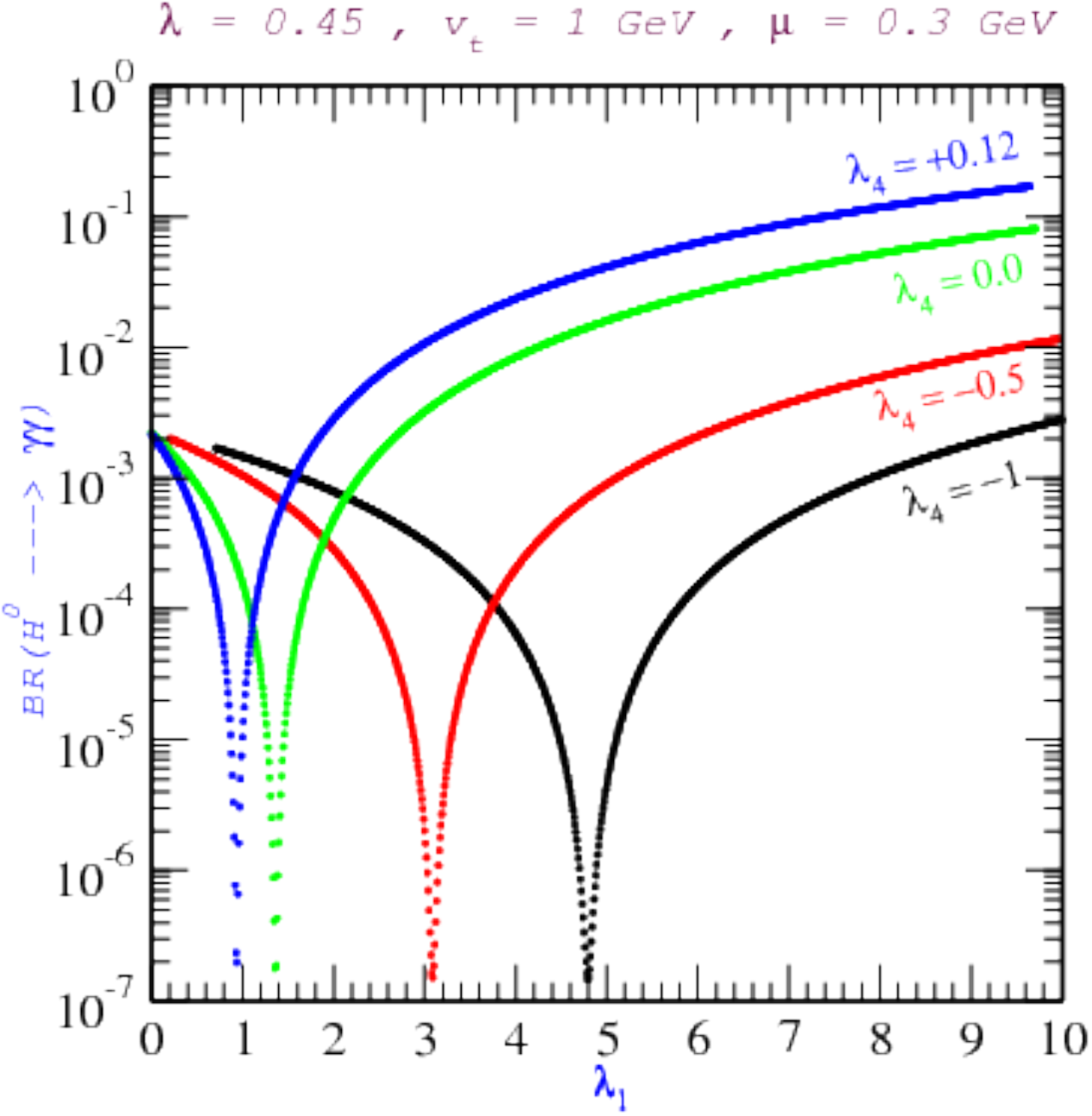}}
  \end{tabular}
\caption{The branching ratios for $\mathcal{H}\rightarrow \gamma\gamma$ as 
a function of $\lambda_1$ for various values of $\lambda_4$ 
with $\lambda=0.45, \lambda_3=2\lambda_2=0.2$ and $v_t=1$ GeV; 
left panel: $\mu=1$~GeV,
$h^0$ is SM-like and $m_{h^0}=114$--$115$~GeV; right panel: $\mu=0.3$~GeV, 
$H^0$ is SM-like and $m_{H^0}=115$--$123$~GeV.}
\label{fig1}
\end{figure}

We show in Fig.~\ref{fig1} the branching ratio for the CP-even
Higgs bosons decays into two photons as a function of $\lambda_1$, illustrated 
for several values of $\lambda_4$ and $\lambda=0.45$, $v_t=1$ GeV.  In the left panel we take 
$\mu=1$~GeV, implying that the lightest CP-even state $h^0$ carries $99$\% of the SM-like Higgs
component, with an essentially fixed mass 
$m_{h^0}\approx 114$--$115$~GeV over the full range of values considered for  $\lambda_1$ and $\lambda_4$.
In the right panel, where $\mu=0.3$~GeV, the heaviest CP-even state $H^0$ carries most of the SM-like Higgs 
component [$\sim 90$\% for $\lambda_1 \lsim 3$] with a mass more sensitive to 
the $\lambda_1$ and $\lambda_4$ couplings, $m_{H^0}\approx 115$--$123$ GeV.\footnote{In the latter case
one has to be cautious in the range $\lambda_1 \lsim 4$--$10$ where $H^0$ carries only
$75$--$85$\% of the SM-like component. The effects of the lighter state $h^0$ with a reduced coupling
to the SM particles and a mass between $102$--$110$~GeV,
should then be included in the estimate of the overall diphoton cross-section.}

As can be seen from the plots, ${\rm Br}(\mathcal{H}\to \gamma\gamma)$ is very close to the SM
prediction [$\approx 2\times 10^{-3}$] for small values of $\lambda_1$, irrespective of the values of $\lambda_4$.
Indeed in this region the diphoton decay is dominated by the SM contributions, the $H^{\pm \pm}$ 
contribution being shutdown for vanishing  $\lambda_1$, cf. Eq.(\ref{eq:gcalHHpp}), while 
the sensitivity to $\lambda_4$ in the $H^\pm$ contribution, Eqs.~(\ref{eq:redgcalHHp}, \ref{eq:gcalHHp}),
is suppressed by a large $m_{H^\pm}$ mass, $m_{H^\pm}\approx 164$--$237$~GeV for $-1 <  \lambda_4 <1$.
Increasing $\lambda_1$ (for fixed $\lambda_4$) enhances the $g_{\mathcal{H}H^\pm H^\mp}$ and 
$g_{\mathcal{H}H^{\pm\pm} H^{\mp\mp}}$ couplings. The destructive
interference, already noted in section 3, between the SM loop contributions and those of
$H^\pm$ and $H^{\pm \pm}$ becomes then more and more pronounced.
The leading DTHM effect is mainly from the $H^{\pm \pm}$ contribution, the latter
being enhanced with respect to $H^\pm$ by a factor $4$ due to the doubled electric charge,
but also  due to a smaller mass than the latter in some parts of the parameter space,  
$m_{H^{\pm \pm}} \approx 110$--$266$~GeV.  It is easy to see from Eqs.~(\ref{eq:DTHM-h2gaga},
\ref{eq:redgcalHHpp} -- \ref{eq:gcalHHp})
that the amplitude for $\mathcal{H}\to \gamma\gamma$ is essentially linear in $\lambda_1$,
since $m_{H^\pm}$ and $m_{H^{\pm\pm}}$, Eqs.~(\ref{eq:mHpm}, \ref{eq:mHpmpm}), do not depend on 
$\lambda_1$ while the dependence on this coupling through $m_{\mathcal{H}}$ is screened by the 
mild behavior of the scalar functions  $A_{0, 1/2, 1}^{{\mathcal{H}}}$. Furthermore,
the latter functions remain real-valued in the considered domain of Higgs masses. 
There exit thus necessarily values of $\lambda_1$ where the effect of the destructive interference
is maximized leading to a tremendous reduction of $\Gamma(\mathcal{H}\to \gamma\gamma)$.
Since all the other decay channels remain SM-like, the same reduction occurs
for ${\rm Br}(\mathcal{H}\to \gamma\gamma)$. 
 The different dips seen in Fig.~\ref{fig1} are due to such a severe 
cancellation between SM loops and $H^\pm$ and $H^{\pm\pm}$ loops, and they occur for
$\lambda_1$ values within the allowed unitarity \& BFB regions. Increasing $\lambda_1$
beyond the dip values, the contributions of 
$H^{\pm\pm}$ and $H^{\pm}$  become bigger than the SM contributions and eventually
come to largely dominate for sufficiently large $\lambda_1$. There is however another
interesting effect when $\lambda_4$ increases. Of course the locations of the dips depend also on
the values of $\lambda_4$, moving them to lower values of  $\lambda_1$
 for larger $\lambda_4$. Thus, for larger $\lambda_4$, there is place, within the considered
range of $\lambda_1$, for a significant increase of ${\rm Br}(\mathcal{H}\to \gamma\gamma)$ by even 
more than one order of magnitude with respect to the SM prediction. 
This spectacular enhancement is due to the fact that larger $\lambda_4$
leads to smaller $H^{\pm\pm}$ and $H^\pm$ which can efficiently boost the reduced
couplings that scale like the inverse second power of these masses. 
For instance varying $\lambda_4$ between $-1$ and $1$ in the left panel case,
decreases $H^{\pm\pm}$ from $266$ to $110$ GeV, while varying it from $-1$ to $0$
in the right panel case decreases $H^{\pm\pm}$ from $205$ to $112$ GeV.
In both cases we see {\rm Br}($\mathcal{H}\to \gamma\gamma$) rising by 2 orders of magnitude
with respect to the SM value.

\begin{figure}[!h]
\hspace{-0.9cm}
\includegraphics[width=2.82in,angle=0]{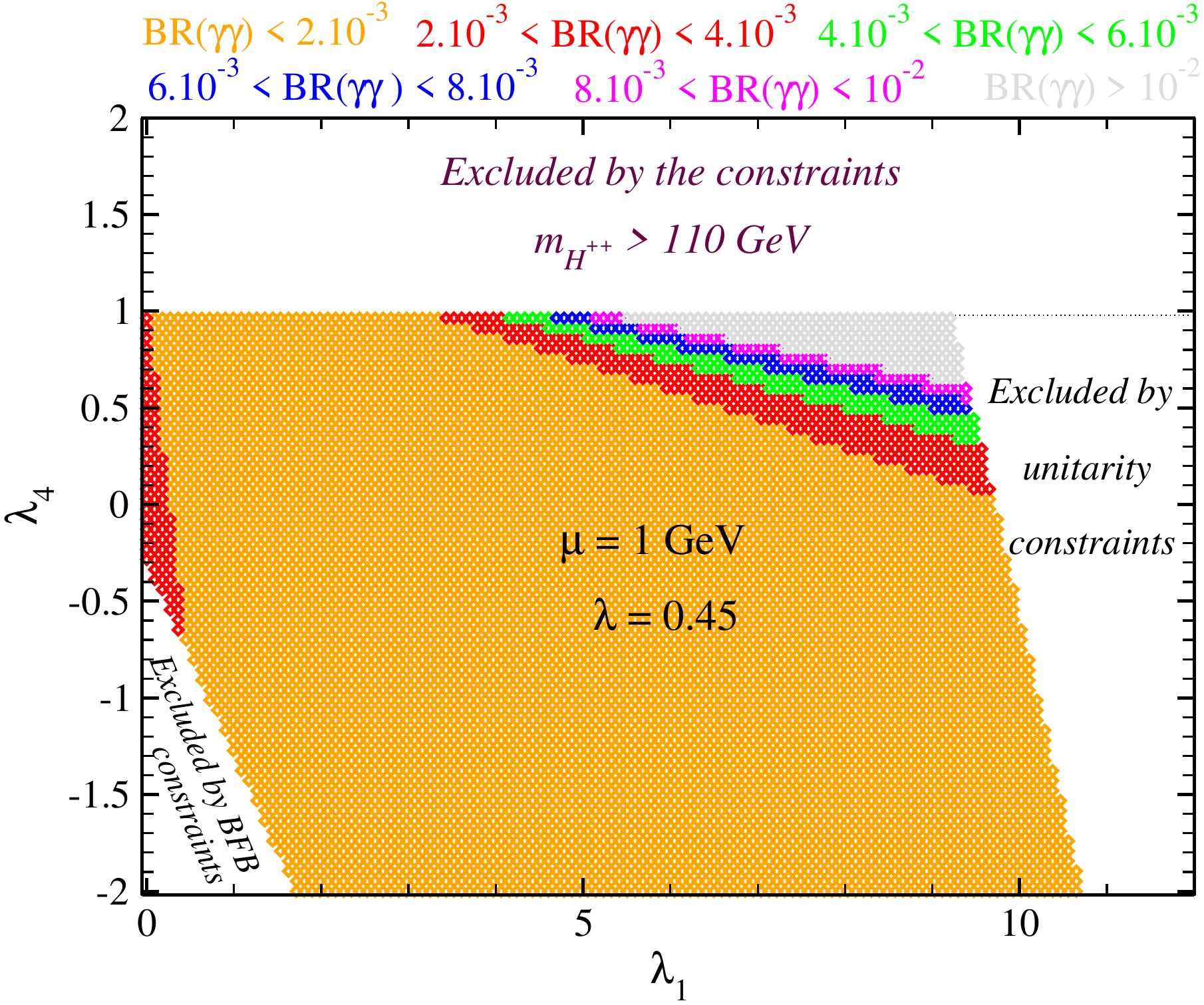}
\includegraphics[width=2.82in,angle=0]{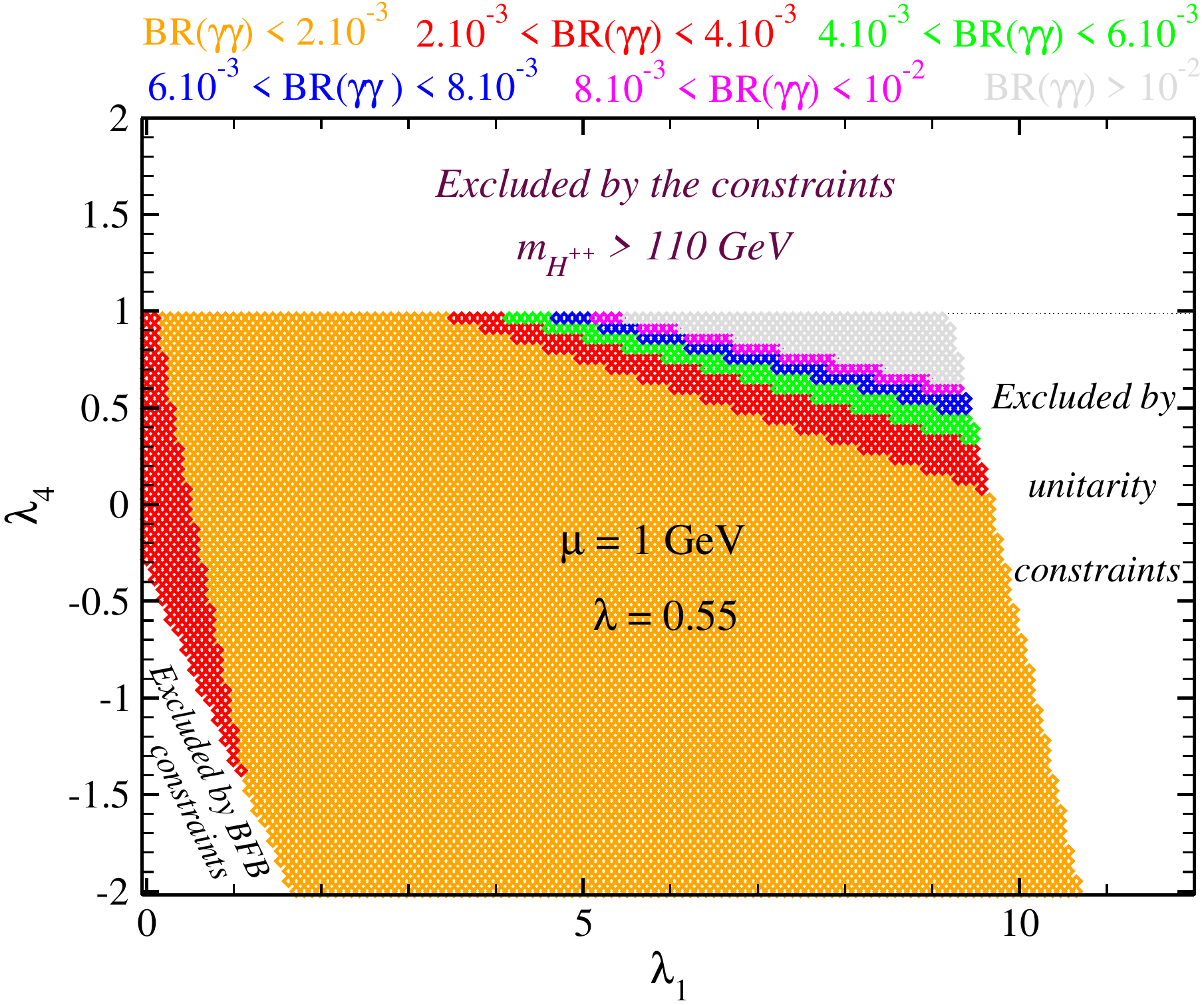}
\caption{Scatter plot  in the $(\lambda_1,\lambda_4)$ plane showing 
the branching ratios for $\mathcal{H}\rightarrow \gamma\gamma$.
In both panels the SM-like Higgs is $h^0$, with $\lambda=0.45$, $m_{h^0}\approx 115$GeV (left panel)
and $\lambda=0.55$, $m_{h^0} \approx 127$GeV (right panel);  
$\lambda_3=2\lambda_2=0.2$ and $\mu=v_t=1$ GeV.}
\label{fig2}
\end{figure}
 
In Fig.~\ref{fig2} we show a scatter plot for Br($\mathcal{H}\to \gamma\gamma$) 
in the $(\lambda_1,\lambda_4)$ plane illustrating more generally
the previously discussed behavior, for $m_{\mathcal{H}} = 115$~GeV (left)
and $m_{\mathcal{H}} = 127$~GeV (right), imposing 
unitarity and BFB constraints as well as the lower bounds $m_{H^\pm} \gtrsim 80$~GeV  
and $m_{H^{\pm \pm}} \gtrsim 110$~GeV on the (doubly-)charged Higgs masses.
One retrieves the gradual enhancement of Br($ \mathcal{H}\to \gamma\gamma$) 
in the regions with large and positive $\lambda_{1,4}$. The largest region (in yellow) corresponding to  
${\rm Br}(\mathcal{H}\to \gamma\gamma)\la 2\times 10^{-3}$ encompasses three cases: --the SM dominates 
--complete cancellation between SM and $H^\pm$, $H^{\pm\pm}$ loops --$H^\pm$, $H^{\pm\pm}$ loops dominate
but still leading to a SM-like branching ratio.

\begin{figure}[!h]
\hspace{-1.1cm}
  \begin{tabular}{cc}
    \resizebox{72mm}{!}{\includegraphics[scale=0.9]{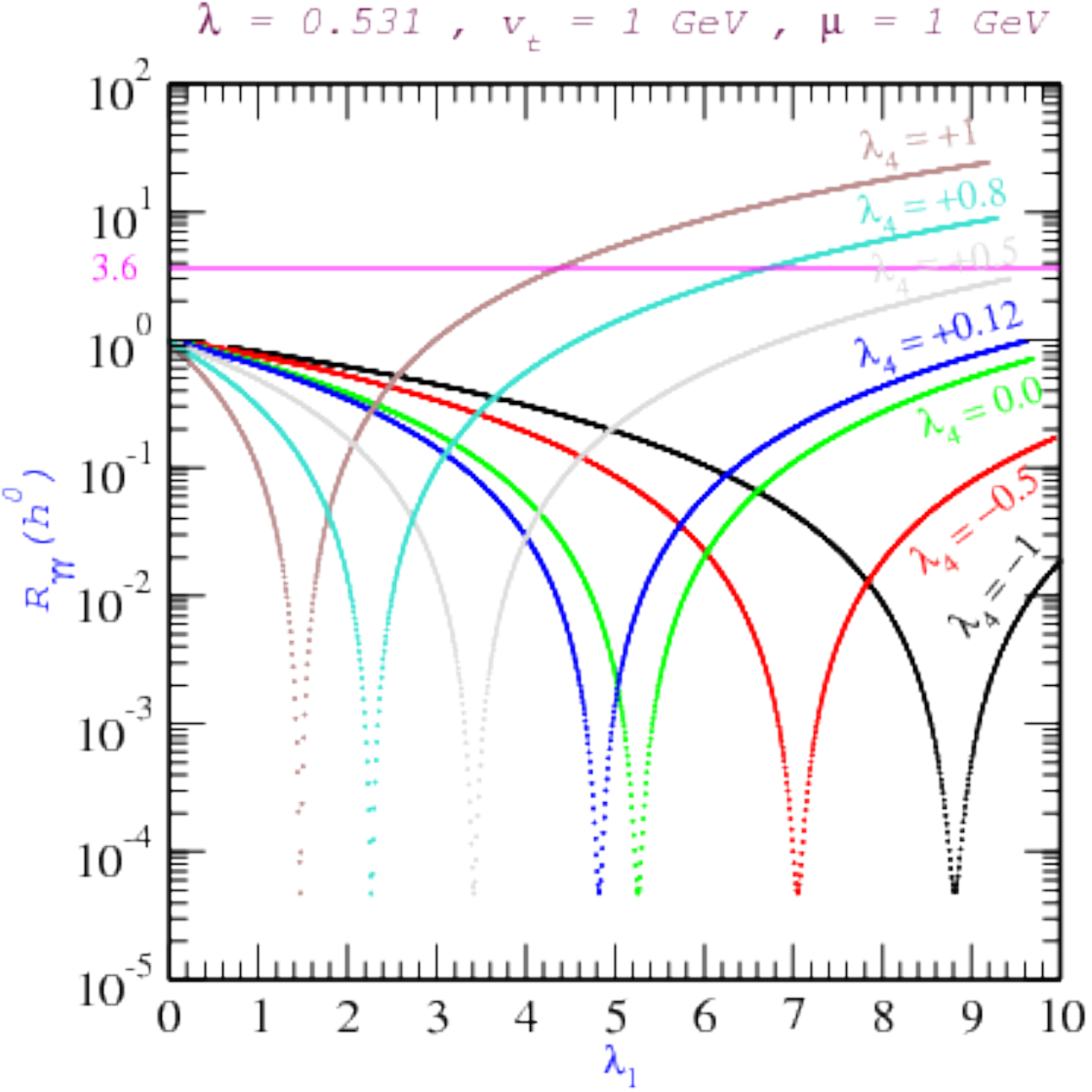}} &
    \resizebox{72mm}{!}{\includegraphics[scale=0.9]{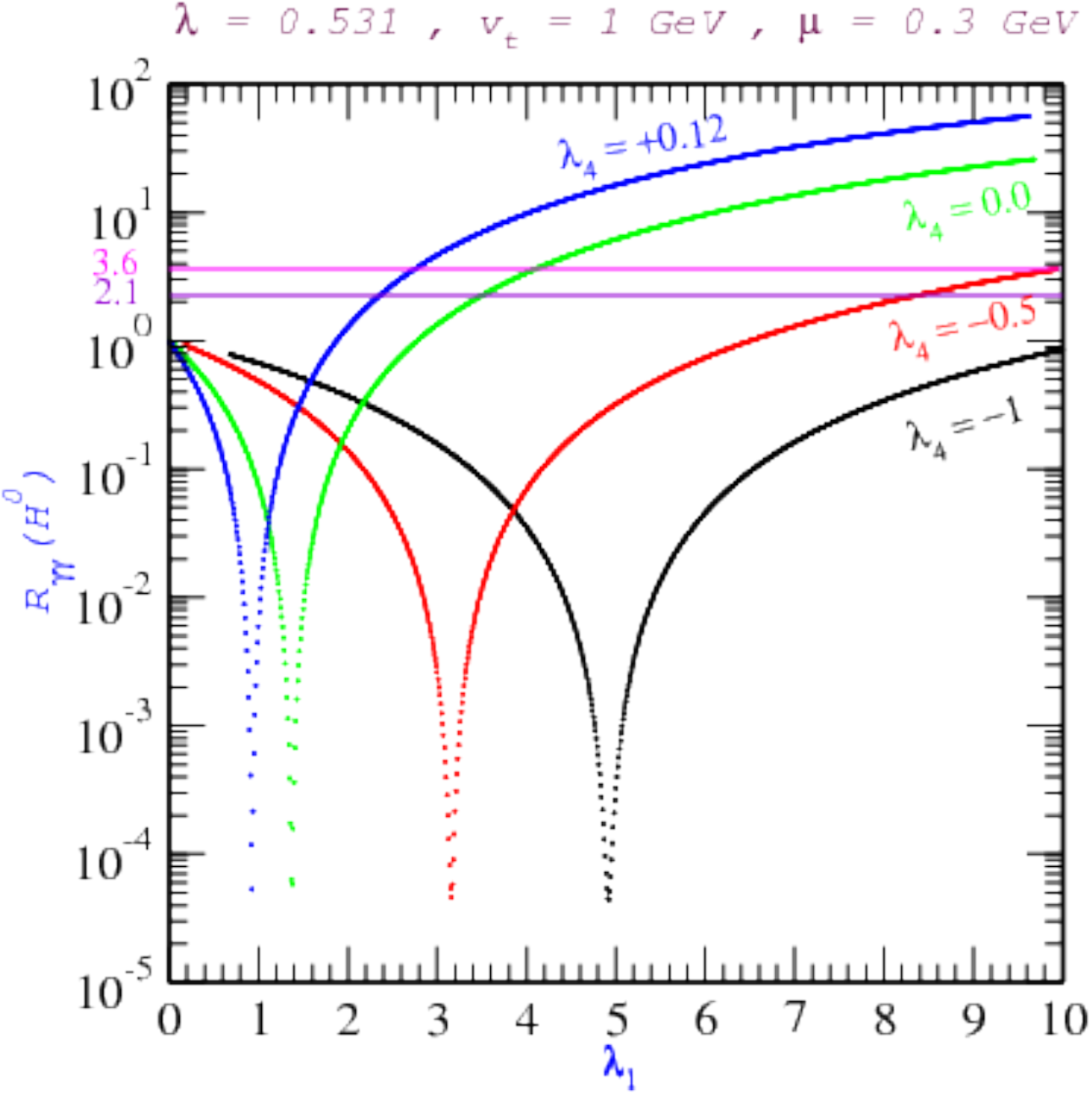}}
  \end{tabular}
\caption{The ratio $R_{\gamma\gamma}$ as a function of 
$\lambda_1$ for various values of $\lambda_4$, 
with $\lambda=0.53, \lambda_3=2\lambda_2=0.2$ and $v_t=1$ GeV; left panel: 
$\mu=1$~GeV,
$h^0$ is SM-like and $m_{h^0}=124$--$125$~GeV; right panel: $\mu=0.3$~GeV, 
$H^0$ is SM-like and $m_{H^0}=125$--$129$~GeV. The horizontal lines in 
both panels indicate the ATLAS exclusion limits \cite{ATLAS-CONF-2011-161} for 
$m_{h^0}=125$~GeV (left) and $m_{H^0}=125$ and $129$ GeV (right).}
\label{fig3}
\end{figure}

\begin{figure}[!ht]
\hspace{-1.1cm}
  \begin{tabular}{cc}
    \resizebox{72mm}{!}{\includegraphics[scale=0.9]{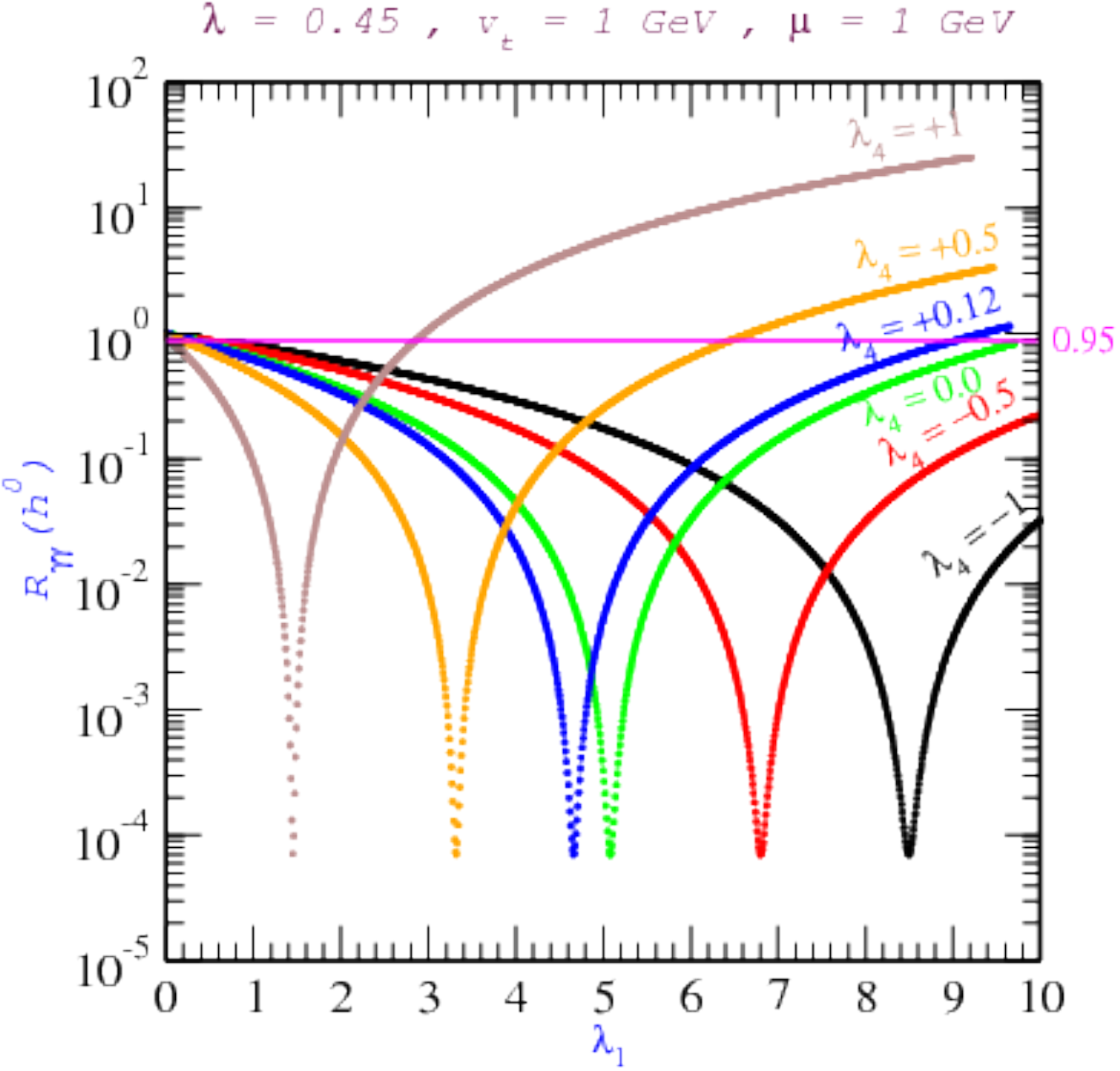}} &
    \resizebox{72mm}{!}{\includegraphics[scale=0.9]{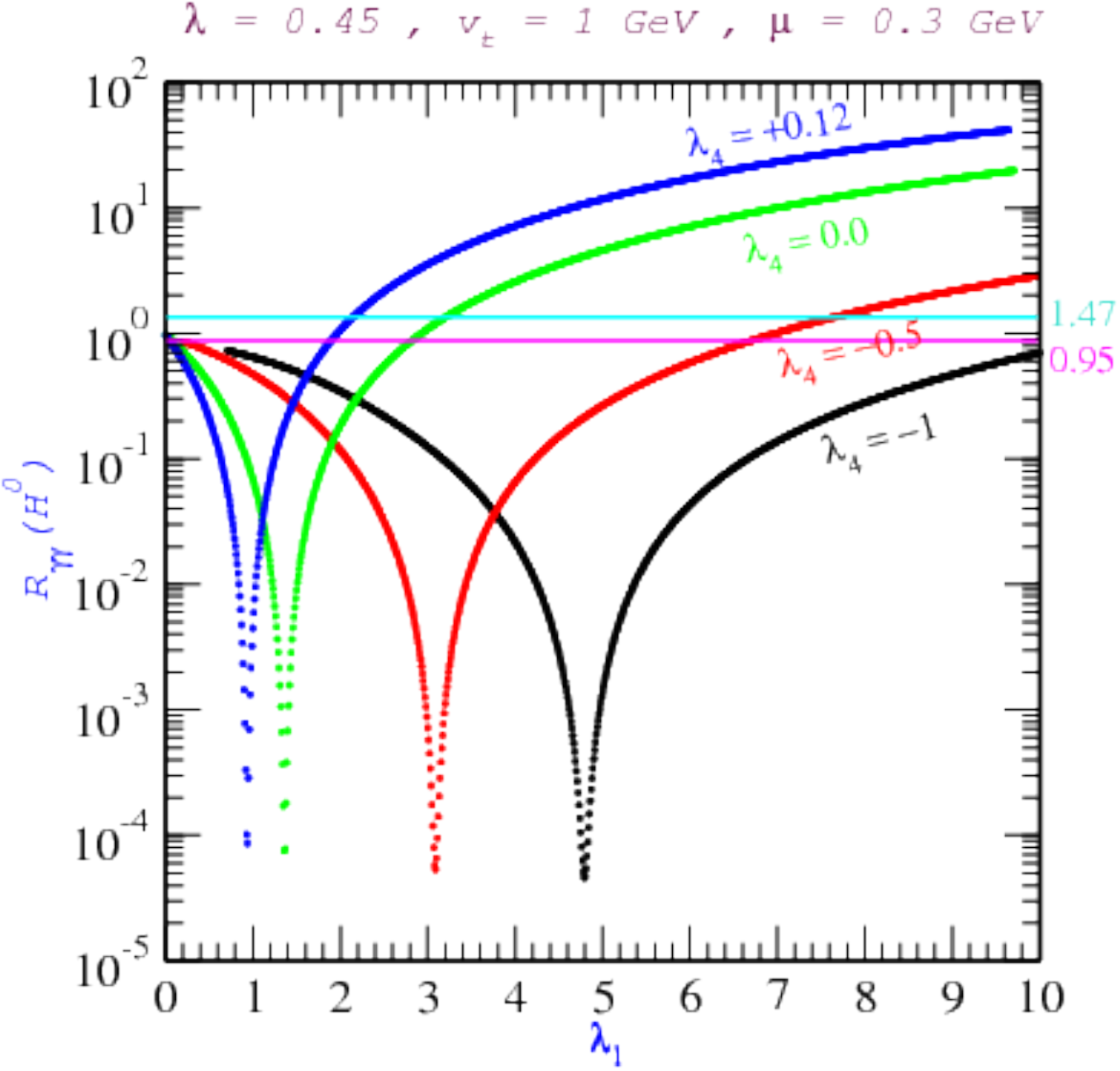}}
  \end{tabular}
\caption{The ratio $R_{\gamma\gamma}$ as a function of 
$\lambda_1$ for various values of $\lambda_4$, (other parameters 
like in Fig.~\ref{fig1}). 
The horizontal lines in both panels indicate the ATLAS exclusion 
limits \cite{ATLAS-CONF-2011-161}
for $m_{h^0}=115$~GeV (left) and $m_{H^0}=115$  and $122.5$ GeV (right).}
\label{fig4}
\end{figure}

In Figs.~\ref{fig3}, \ref{fig4} we illustrate the effects directly in terms of the ratio 
$R_{\gamma \gamma} \approx \sigma^{\gamma \gamma}/\sigma_{\rm SM}^{\gamma \gamma}$ defined
in Eq.(\ref{eq:Rgg}), for 
benchmark Higgs masses.  
We also show on the plots the present experimental exclusion limits corresponding to these masses, taken
from \cite{ATLAS-CONF-2011-161}. As can be seen from Fig.\ref{fig3}, one can easily accommodate,
for $m_{\mathcal{H}} \approx 125 {\rm GeV}$,
a SM cross-section, $R_{\gamma \gamma}(m_{\mathcal{H}}=125 {\rm GeV})=1$, or a cross-section in excess
of the SM, e.g. $R_{\gamma \gamma}(m_{\mathcal{H}}=125 {\rm GeV}) \sim 3$--$4$, for values of 
$\lambda_1, \lambda_4$ within the theoretically allowed region, fulfilling as well the present 
experimental bound $m_{H^\pm} \gtrsim 80$~GeV and the moderate bound $m_{H^{\pm\pm}} \gtrsim 110$~GeV
as discussed previously. The excess reported by ATLAS and CMS in the diphoton channel can be readily 
interpreted in this context. However, one should keep in mind that all other channels remain
SM-like, so that the milder excess observed in $WW^*$ and $ZZ^*$ should disappear with higher
statistics in this scenario. This holds independently of which of the two states, $h^0$ or $H^0$, 
is playing the role of the SM-like Higgs. 

We comment now on another scenario, in case the reported excess around 
$m_{\mathcal{H}} \approx 125$~GeV would not stand the future accumulated statistics.
Fig.\ref{fig4} shows the $R_{\gamma \gamma}$ ratio corresponding to the case of
Fig.~\ref{fig1} with $m_{\mathcal{H}}$ close to $115$~GeV. The large deficit 
for $R_{\gamma \gamma}$ in parts of the $(\lambda_1, \lambda_4)$
parameter space opens up an unusual possibility: the exclusion of a SM-like Higgs, such as the
one reported by ATLAS in the $114$--$115$~GeV range, does not exclude the LEP events as being
real SM-like Higgs events in the same mass range! 
This is a direct consequence of the fact that in the model we consider, even a tremendous reduction in 
$\sigma^{\gamma \gamma}=\sigma^{\mathcal{H}} \times {\rm Br}(\mathcal{H} \to \gamma \gamma)$ 
leaves all other channels, and in particular the LEP relevant cross-section $\sigma(e^+ e^- \to
Z \mathcal{H})$ essentially identical to that of the SM.

\begin{figure}[!h]
\centering
\hspace{-1.cm}\includegraphics[width=2.8in,angle=0]{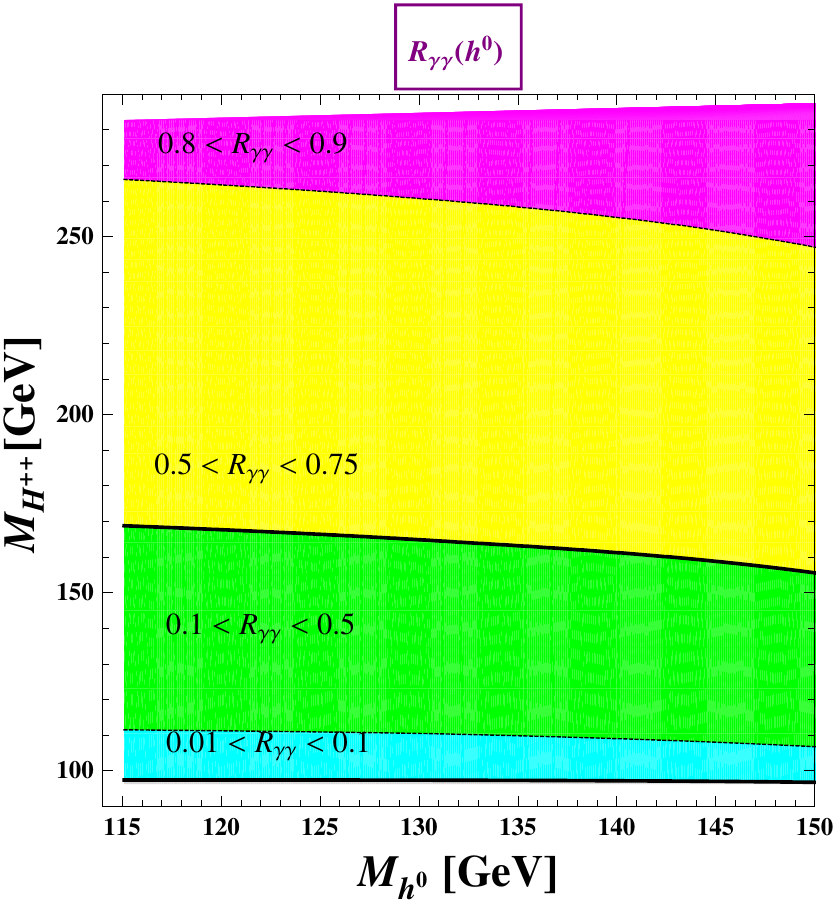}
\includegraphics[width=2.8in,angle=0]{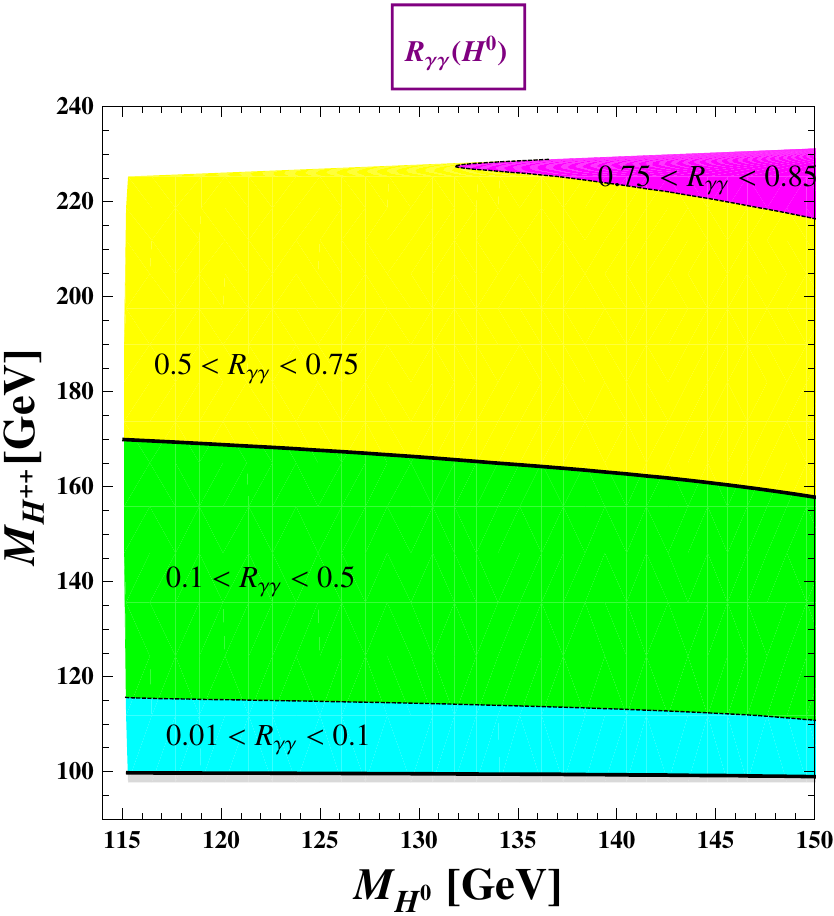}
\caption{Scatter plots in the ($m_{h^0}, m_{H^{\pm \pm}}$) and 
($m_{H^0}, m_{H^{\pm \pm}}$) planes, 
with $h^0$ SM-like ($\mu= 1$GeV, left panel) and $H^0$ SM-like 
($\mu=.3$ GeV, right panel), showing
domains of  $R_{\gamma\gamma}$ values. We scan in the domain 
$.45 < \lambda < 1, -5 < \lambda_4 < 3$ with 
$\lambda_1=1$, $\lambda_3=2\lambda_2=0.2$ and $v_t=1$ GeV,
consistent with the unitarity and BFB constraints and
requiring $m_{H^\pm} \gtrsim 80$~GeV.}
\label{fig5}
\end{figure}
\begin{figure}[]
\centering\hspace{-1.cm}\includegraphics[width=2.8in,angle=0]{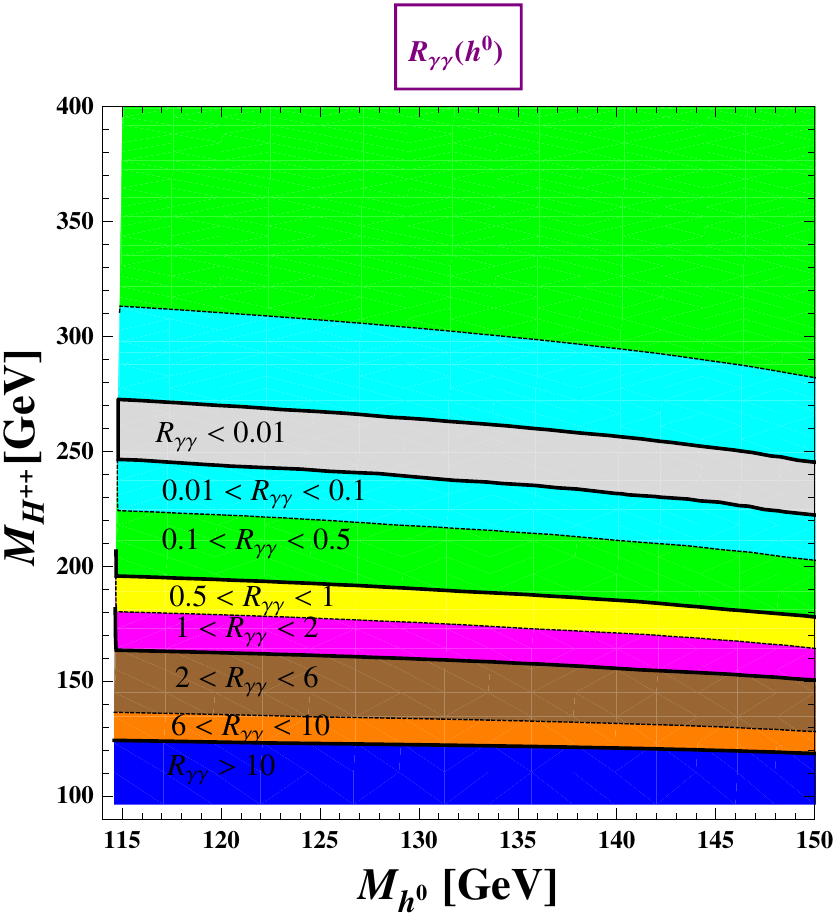}
\includegraphics[width=2.8in,angle=0]{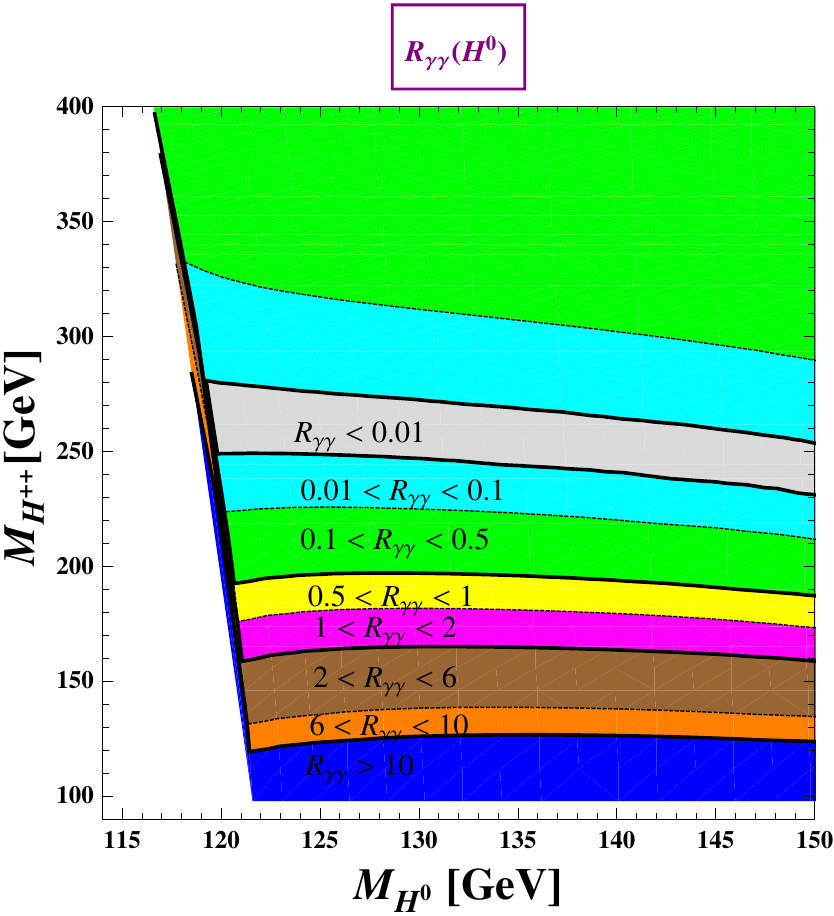}
\includegraphics[width=2.8in,angle=0]{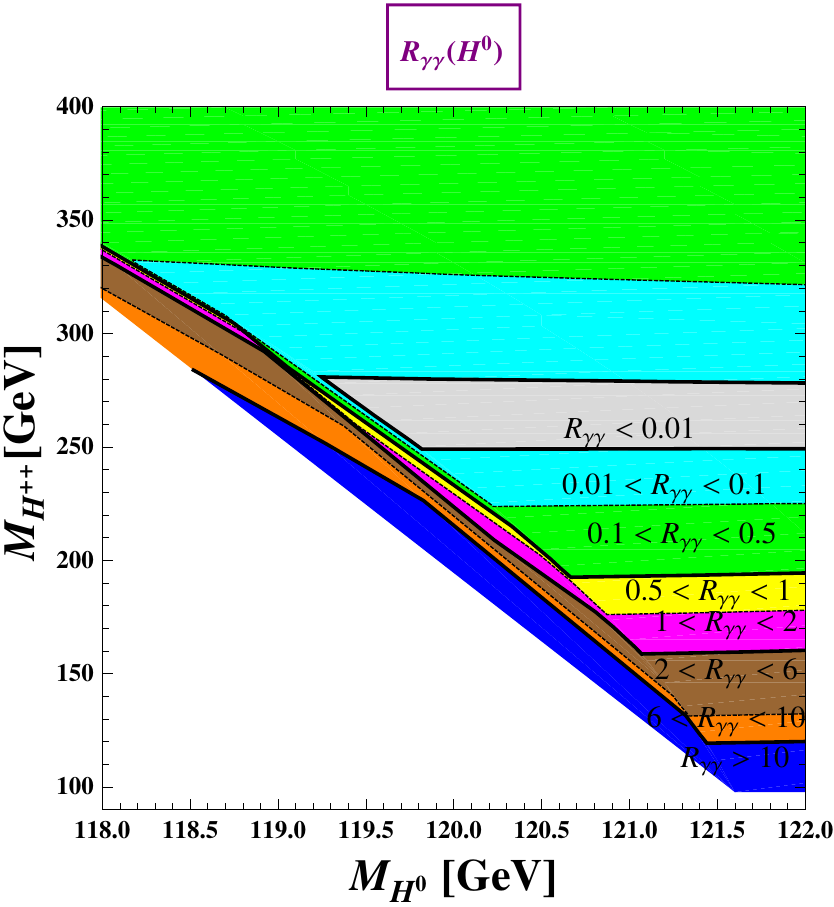}
\caption{Scatter plots in the ($m_{h^0}, m_{H^{\pm \pm}}$) and ($m_{H^0}, m_{H^{\pm \pm}}$) planes, 
with $h^0$ SM-like ($\mu= 1$GeV, upper left panel) and $H^0$ SM-like ($\mu=.3$ GeV, upper right panel),
showing domains of  $R_{\gamma\gamma}$ values. The lower plot zooms on the distinctive features in the
$H^0$ SM-like case. The scanned domains are as in Fig.~\ref{fig5} but with 
$\lambda_1=8$.}
\label{fig6}
\end{figure}

Last but not least,  exclusion limits or a signal in the diphoton channel can be translated into
constraints on the masses of $H^{\pm \pm}$ and $H^\pm$. We show in Figs.~\ref{fig5} and \ref{fig6}
the correlation between $m_{\mathcal{H}}$ and $m_{H^{\pm \pm}}$ for different ranges of
$R_{\gamma \gamma}$. Obviously, the main dependence on $m_{\mathcal{H}}$ drops out in the
ratio $R_{\gamma \gamma}$ whence the almost horizontal bands in the plots. There remains however
small correlations which are due to the model-dependent relations between the (doubly-)charged
and neutral Higgs masses that can even be magnified in the regime of $H^0$ SM-like, albeit
in a very small mass region  (see bottom panel of \ref{fig6}). The sensitivity to the 
coupling $\lambda_1$ can be seen by comparing Figs.~\ref{fig5} and \ref{fig6}.
For low values of $\lambda_1$ as in Fig.~\ref{fig5}, the ratio $R_{\gamma \gamma}$ remains below $1$ even for
increasing $H^{\pm \pm}$ and $H^\pm$ masses.  The reason is that these masses become large when
$\lambda_4$ is large (and negative)  for which the loop contribution of $H^\pm$ does
not vanish, as can be easily seen from Eqs.~(\ref{eq:mHpm}, \ref{eq:redgcalHHp}, \ref{eq:gcalHHp}).

In contrast, we see that
for the parameter set of Fig.~\ref{fig6}, $R_{\gamma \gamma}$ can take SM-like values 
for $m_{H^{\pm \pm}}$ of order $180$~GeV, while an excess of $2$ to $6$ can be achieved 
for $m_{H^{\pm \pm}} \approx 130$--$160$~GeV, and a  
deficit  in $R_{\gamma \gamma}$, down to $2$ orders of magnitude, for $m_{H^{\pm \pm}}$
between $200$ and $300$~GeV. Increasing $m_{H^{\pm \pm}}$ (and $m_{H^{\pm}}$) further,
increases $R_{\gamma \gamma}$ again, but rather very slowly towards the SM expectation 
as can be seen from the upper green region of the plots.

\section{Conclusions}
\label{sec:conclusion}

The very recent ATLAS and CMS exclusion limits for the search for the Higgs boson, clearly
indicate that if such a light SM-like state  exists, it should be somewhere in the region
between $114.4$~(LEP) and $130$~GeV. The diphoton channel is thus expected to play
a crucial role in the near future data analyses, eventually confirming the not yet
statistically significant excess around $125$~GeV. In this paper we have shown
that the diphoton channel can be interpreted in a peculiar way in the context of the
Type II Seesaw model, provided that the full Higgs sector of the model 
lies below the TeV scale. While there is always in this model a neutral Higgs state coupling
essentially like the SM Higgs, the diphoton channel can be drastically enhanced or reduced by 
several factors with respect to the SM prediction, as a result of the loop effects of the 
doubly-charged (and charged) Higgs states, while all the other relevant decay (as well as 
production) channels remain at their SM level. Theoretically consistent domains of the parameter
space in the small $\mu$ regime can thus account either for an excess or for a deficit or even for a SM 
value of the diphoton cross-section, making the model hard to rule out on the basis of
the neutral Higgs observables alone.  In particular, the exclusion of a mass region
through the diphoton channel does not exclude SM-like Higgs events in the other channels
(including the LEP $ZH$ channel) for the same mass region. Rather, it can be re-interpreted in
terms of bounds on the masses of the doubly-charged (and charged) Higgs states of the model.
The experimental search for such light doubly-charged states through their decay into (off-shell)
W bosons is a crucial test of the model while the present bounds based on same-sign di-lepton decays 
do not necessarily apply in our scenario.

\section*{Acknowledgments}
We would like to thank Goran Senjanovic, Dirk Zerwas and Johann Collot for very useful discussions. 
A.A. and M.C. would like to thank NCTS-Taiwan for partial support, and 
Chuang-Huan Chen and Hsiang-nan Li for discussions and hospitalty at NCKU and 
Academia Sinica. This work was supported by Programme Hubert Curien, 
Volubilis, AI n0: MA/08/186. 
We also acknowledge the ICTP-IAEA Training Educational Program for partial
support, as well as the LIA (International Laboratory for Collider Physics-ILCP).


\bibliographystyle{unsrt}
\bibliography{references-triplet}

\end{document}